\documentclass{jkas}

\def\year{2019} % publication year
\def\month{June} % publication month
\def\volume{52} % journal volume
\def\issue{3} % journal issue
\def\beginpage{71} % first page of article
\def\endpage{82} % last page of article
\def\received{January 25, 2019} % date paper was received by JKAS
\def\accepted{May 21, 2019} % date of acceptance
\date{Received \received; accepted \accepted}

\usepackage{float}
\usepackage{adjustbox}
\usepackage{enumerate}
\usepackage{flushend}
\usepackage{microtype}

\usepackage[colorlinks,
            pdftex,
            breaklinks,
            linkcolor=blue,
            urlcolor=blue,
            anchorcolor=blue,
            menucolor=blue,
            citecolor=blue]{hyperref}

\renewcommand{\lg}{\ensuremath{\log_{10}}}
\newcommand{\pV}{\ensuremath{p_\mathrm{V}}}
\newcommand{\HV}{\ensuremath{H_\mathrm{V}}}
\newcommand{\HVa}{\ensuremath{H_\mathrm{V}(\alpha)}}
\newcommand{\h}{\ensuremath{\mathrm{h}}}
\newcommand{\g}{\ensuremath{\mathrm{g}}}

\newcommand{\proj}{\ensuremath{\mathrm{proj}}}

\newcommand{\pVval}{\ensuremath{0.185^{+0.045}_{-0.039}}}
\newcommand{\Aval}{\ensuremath{0.104^{+0.050}_{-0.034}}}
\newcommand{\faval}{\ensuremath{1.708^{+1.285}_{-0.740}}}
\newcommand{\fbval}{\ensuremath{0.018^{+0.019}_{-0.019}}}

\newcommand{\si}[1]{\ensuremath{ \mathrm{\,\mathrm{#1}} }}
\newcommand{\arcsec}{\ensuremath{^{\prime\prime}}}

\newcommand{\degr}{\ensuremath{^{\circ}}}

\newcommand{\pdv}[2]{\ensuremath{\frac{\partial #1}{\partial #2}}}

\newcommand{\tablefootmark}[1]{\ensuremath{\mathrm{^{#1}}}}

\newcommand*\aap{A\&A}

\newcommand*\aj{AJ}

\newcommand*\apj{ApJ}

\newcommand*\icarus{Icarus}

\newcommand*\mnras{MNRAS}

\newcommand*\pasj{PASJ}
\newcommand*\pasp{PASP}

\newcommand*\planss{Planet.~Space~Sci.}

\def\arxivprefixesep{:}

\newcommand{\eprint}[2][]{
{\tt\if!#1!#2\else#1\arxivprefixesep\ignorespaces#2\fi}%
}

\title{The Geometric Albedo of (4179) Toutatis Estimated from\\ KMTNet DEEP-South Observations}

\author[1]{Yoonsoo~P.~Bach}
\author[1]{Masateru~Ishiguro}
\author[1]{Sunho~Jin}
\author[2]{Hongu~Yang}
\author[2]{Hong-Kyu~Moon}
\author[2]{Young-Jun~Choi}
\author[2]{Youngmin~JeongAhn}
\author[2]{Myung-Jin~Kim}
\author[2]{SungWon~Kwak}

\affil[1]{Department of Physics and Astronomy, Seoul National University, 1 Gwanak, Seoul 08826, Korea; \email{ishiguro@astro.snu.ac.kr}}
\affil[2]{Korea Astronomy and Space Science Institute, 776, Daedeok-daero, Yuseong-gu, Daejeon 34055, Korea}

\def\corrauthor{M. Ishiguro}
\def\runningauthor{Bach et al.}
\def\runningtitle{The Geometric Albedo of (4179) Toutatis from KMTNet DEEP-South Observations}
\def\keywords{minor planets, asteroids: general --- minor planets, asteroids: individual: (4179) Toutatis}

\def\abstracttext{
We derive the geometric albedo of a near-Earth asteroid, (4179) Toutatis, to investigate its surface physical conditions. The asteroid has been studied rigorously not only via ground-based photometric, spectrometric, polarimetric, and radar observations but also via \textit{in situ} observation by the Chinese Chang'e-2 space probe; however, its geometric albedo is not well understood.
We conducted V-band photometric observations when the asteroid was at opposition in April 2018 using the three telescopes in the southern hemisphere that compose the Korea Microlensing Telescope Network (KMTNet). The observed time-variable cross section was corrected using the radar shape model.
We find that Toutatis has a geometric albedo $\pV = \pVval $, which is typical of S-type asteroids. We compare the geometric albedo with archival polarimetric data and further find that the polarimetric slope--albedo law provides a reliable estimate for the albedo of this S-type asteroid. The thermal infrared observation also produced similar results if the size of the asteroid is updated to match the results from Chang'e-2. We conjecture that the surface of Toutatis is covered with grains smaller than that of the near-Sun asteroids including (1566) Icarus and (3200) Phaethon.
}

\begin{document}
\jkashead %% set title, authors, abstract, etc.

\section{Introduction\label{sec:intro}}

The geometric albedo is one of the most fundamental observed quantities for characterizing the physical properties and composition on asteroids. It provides information on the surface composition, degree of space weathering, and  physical conditions such as the surface particle size if polarimetric data at large phase angles (Sun--asteroid--observer angle) are available \citep{1992Icar...99..468S,1998Icar..136...69D}.
Because the absolute photometric magnitude (at the opposite direction from the Sun viewed at unit heliocentric and observer distances) is related to the product of the geometric albedo and cross section of the asteroid, the accuracy of the size determination is a critical factor for deriving the geometric albedo in most cases. The sizes of asteroids have been investigated thoroughly via various techniques, namely, direct imaging of asteroids with cameras onboard spacecraft, delay-Doppler imaging \citep{1993RvMP...65.1235O}, thermal--infrared observations \citep{2002AJ....123.1056T,2011ApJ...731...53M,2011PASJ...63.1117U}, and occultation \citep{2006Icar..184..211S}; nevertheless, absolute magnitudes are rarely studied because of the scarcity of observational opportunities at opposition. In addition, magnitude variations due to an asteroid's rotation have not been considered in many cases for the derivation, making the albedo values less certain. For these reasons, the number of reliable albedo estimates for asteroids is limited.

Here, we examine the geometric albedo of an Apollo-type near-Earth asteroid (NEA), (4179) Toutatis. The asteroid has been thoroughly studied by various observational techniques (i.e., photometry, spectroscopy, polarimetry, radar imaging, and {\it in situ} observation). It is classified into S-type \citep{1994P&SS...42..327L,1994Icar..111..468H} or, more strictly, $ \mathrm{S_k} $-type \citep{2002Icar..158..146B} and is possibly composed of undifferentiated L-chondrites \citep{2012Icar..221.1177R}. Its physical dimensions are $ 4.6 \si{km} \times 2.4 \si{km} \times 1.9 \si{km} $ according to radar observations \citep{1995Sci...270...84H}. Toutatis is in an extremely slow, non-principal-axis rotational state \citep{1995Sci...270...80O, 2013AJ....146...95T} having rotation and precession periods of $ \sim 5.38 \si{days} $ and $ \sim 7.40 \si{days} $, respectively \citep{ZhaoY+2015MNRAS_Toutatis_spin_Change2}. An accurate shape model was established by radar delay-Doppler imaging \citep{2003Icar..161..346H}.

More recently, in December 2012, the Chinese spacecraft Chang'e-2 captured images of this asteroid with a maximum resolution better than 3\si{m} during the flyby \citep{2013NatSR...3E3411H}. The study revealed, e.g., a previously unknown $ \sim 805 \si{m} $ basin at the end of the body, lineaments over the surface especially near the rim of it, and a sharp connection silhouette in the neck, as well as providing valuable information on the asteroid's size ($ 4.75 \times 1.95 \si{km} $ with nominal uncertainty of 10\%). Moreover, more than 50 craters larger than 36\si{m} and 30 boulders larger than 10\si{m} were identified and the chronology of craters and size distribution of boulders were investigated.
From the cumulative boulder size frequency distribution analyses compared with (25143) Itokawa, it has been suggested that Toutatis may have a different preservation state or a diverse formation history \citep{JiangY+2015NatSR_Change2_Boulders}. The shape is considered to have been produced by a low-speed impact between two components \citep{2018MNRAS.478..501H}. The hemispherical albedo was found \citep{ZhaoDF+2016AcASn_Change_hemisphr_albedo} to be 0.2083, 0.1269, and 0.1346 in the R, G, and B bands of the Chang'e-2 CMOS censor, respectively. Moreover, the close-orbit dynamics was used to explore Toutatis' complex gravitational field. The asteroid is also a notable for its extremely chaotic orbit in a 3:1 mean motion resonance with Jupiter and a weak 1:4 resonance with Earth \citep{1993Icar..105..408W}. In addition, the asteroid has a unique potential field in the close orbit \citep{1998Icar..132...53S}. Thus, Toutatis has been central to the developments in NEA research since the 1990s.

Although Toutatis has been studied intensively, its geometric albedo is uncertain by a factor of $ > 3 $, ranging from 0.13 \citep{1995Icar..113..200L} to 0.41 \citep{2017AJ....154..168M}.  In this paper, we aim to establish a better estimate of the albedo value. Taking advantage of the continuous observation capability from KMTNet observatories in three different continents \citep[][see Section \ref{sec: observation}]{KimSL+2016JKAS_KMTNet}, we derived the geometric albedo of the slowly rotating asteroid. Together with the polarimetric properties given in \citet{1995Icar..113..200L, 1997Icar..127..452M, 1997PASJ...49L..31I}, we discuss the physical properties of the surface.

For future reference, Table \ref{tab: notation} lists the some important symbols we used in Sections \ref{sec: observation}, \ref{sec: method}, and Appendix \ref{app: rotation}, their meanings, and their values and units. Most source codes of the software we used for the data reduction and analysis are publicly available (see Appendix \ref{app: online}).

\begin{table*}
  \centering
  \caption{Symbols frequently used in this paper \label{tab: notation}}
  \setlength{\tabcolsep}{19pt}
  \begin{tabular}{llll}
    \toprule
    Category & Symbols & Description & Value and Unit \\  % Table Heading
    \midrule
    Magnitudes
     & $ V_\odot $         & Visual magnitude of the Sun           & $ -26.762 \si{(mag)} $       \\
     & $ V $, $ \Delta V $ & Visual magnitude and its uncertainty  & $ \si{(mag)} $ \\
     & $ \HVa $            & Reduced magnitude                     & $ \si{(mag)} $ \\
     & $ \HV $             & Absolute magnitude ($ \equiv \HV(0) $) & $ \si{(mag)} $ \\
     & $ G $               & Slope parameter                       & ~~ --- \\
    \midrule
    Ephemerides
     & $ r_\h $, $ r_\g $           & Heliocentric/geocentric distance          & $ \si{m} $ or $ \si{au} $ \\
     & $ \alpha $                 & Phase angle                               & ($ \si{\degr} $ or $ \si{rad} $) \\
     & $ t $                      & Time on the target (light-time corrected) & $ \si{s} $ or $ \si{JD} $ \\
     & $ (\lambda,\, \beta) $     & Ecliptic coordinate (longitude, latitude) & ($ \si{\degr} $ or $ \si{rad} $) \\
%     & $ (\lambda_\h,\, \beta_\h)_\mathrm{h.e.} $
%                                  & Heliocentric $ (\lambda, \beta) $         & ($ \si{\degr} $ or $ \si{rad} $) \\
%     & $ (\lambda_\g,\, \beta_\g)_\mathrm{g.e.} $
%                                  & Geocentric $ (\lambda, \beta) $           & ($ \si{\degr} $ or $ \si{rad} $) \\
%    \hline
%    Rotation
%     & $ \spin_1 $, $ \spin_2 $
%                                & Rotational/precessional vector              & ($ \si{\degr} $, $ \si{\degr} $)  \\
%     & $ P_1 $, $ P_2 $           & Rotational/precessional period            & $ \si{s} $ or $ \si{day} $ \\
%     & $ \omega_1 $, $ \omega_2 $ & Rotational/precessional angular speed     & $ \si{s^{-1}} $ ($ \si{rad/s} $) \\
%     & $ \varphi_1 $, $ \varphi_2 $ & Rotational/precessional phase offset    & ($ \mathrm{rad} $) \\
    \midrule
    Physical
     & $ S_\proj $      & Total projected area viewed at $ \alpha = 0 $ & $ \si{m^2} $ \\
    parameters
     & $ D $           & Effective diameter                                   & $ \si{m} $ or $ \si{km} $\\
     & $ \pV $         & Geometric albedo in visual (V) band                  & ~~ --- \\
     & $ A_5 $         & Albedo at the phase angle of $ \alpha = 5 \degr $    & ~~ --- \\
     & $ I $           & The irradiance of the object of interest             & $ \si{W/m^2} $ \\
     & $ F $           & $ I $ of a Lambertian reflector at normal incidence  & $ \si{W/m^2} $ \\
     & $ I/F $         & The radiance factor                                  & ~~ --- \\
    \bottomrule
  \end{tabular}
  \tabnote{Parameters with ``---'' in the last column are dimensionless. The units are usually in SI format unless special units are dominantly used in this work. The units given in parentheses are dimensionless but are preferred to be written out explicitly.}
\end{table*}

\section{Observation and Data Reduction \label{sec: observation}}

\begin{figure}[t!]
  \centering
  \includegraphics[width=0.99\linewidth]{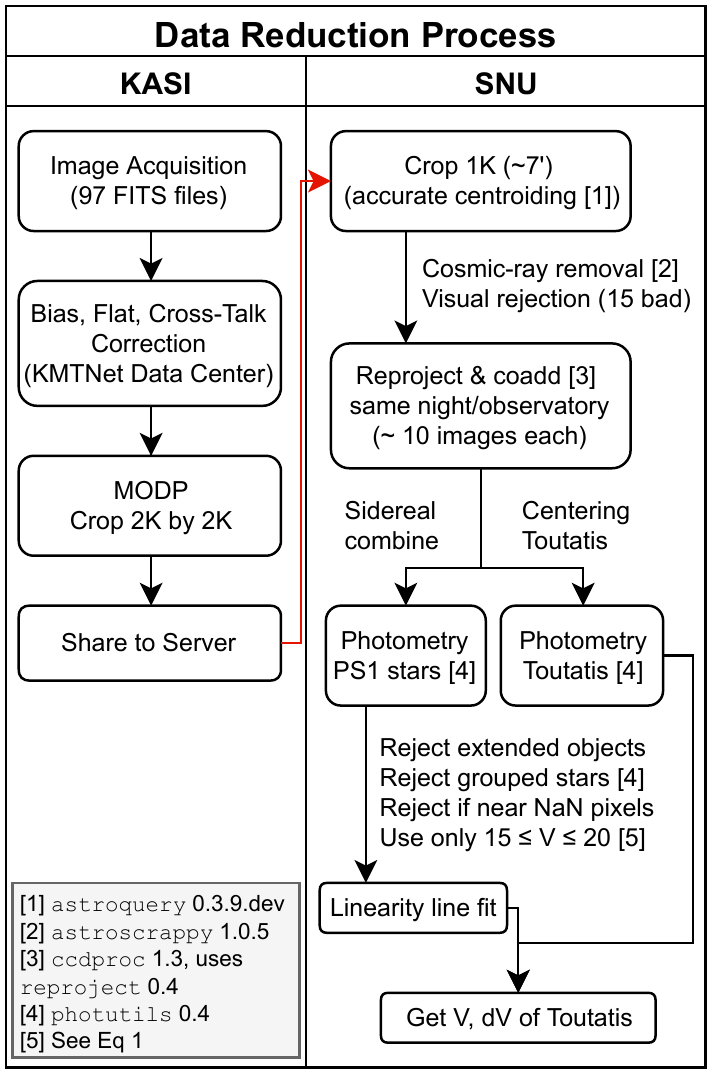}
  \caption{Summary of the data reduction process performed by teams at the Korea Astronomy and Space Science Institute (KASI) and Seoul National University (SNU). Photometry was conducted with publicly available open-source packages indicated at the bottom left (the shadowed box), which includes \texttt{astroquery} \citep{astroquery_v0.3.8}, \texttt{astroscrappy}, \texttt{ccdproc} \citep{ccdproc_v1.3.0}, \texttt{reproject} \citep{reproject_vall}, and \texttt{photutils} \citep{photutils_v0.4}, in addition to the \texttt{astropy} core package \citep{astropy_paper1, astropy_paper2}. }
  \label{fig:toutatisflowchart}
\end{figure}

Our observation journal is provided in Table \ref{table:obslog}. We used three 1.6-m telescopes in the southern hemisphere that together comprise the Korea Microlensing Telescope Network (or KMTNet, \citealt{KimSL+2016JKAS_KMTNet}) within the frame of the DEEP--South project \citep{MoonH-K+2016IAUS_DEEP-South}. Each telescope provides a $ 2 \degr \times 2 \degr $ field of view via identical mosaic CCD cameras with a pixel size of $ 0.4 \arcsec $. The telescopes are located on three continents: at the Cerro Tololo Inter--American Observatory (KMTNet-CTIO) in Chile, the South African Astronomical Observatory (KMTNet-SAAO) in South Africa, and the Siding Spring Observatory (KMTNet-SSO) in Australia. We employed the standard Johnson--Cousins $V$-band filter system. We obtained $\sim$ 10 frames per night at each observatory with individual exposure times of 1 minute. The images were preprocessed by a reduction pipeline for KMTNet data at the KMTNet data center, which performs bias subtraction, flat-fielding, and cross-talk correction. A summary of the data reduction process is given below (see also Figure \ref{fig:toutatisflowchart}).

\begin{table*}[t]
\centering
\caption{Summary of our observations}
\label{table:obslog}
\setlength{\tabcolsep}{4pt}
  \begin{tabular}{cccccccccccc}
    \toprule
    Date\tablefootmark{(a)} & Time in UT\tablefootmark{(a)} & Site\tablefootmark{(b)} & $N$\tablefootmark{(c)} & $r_h$\tablefootmark{(d)} & $ r_\g $\tablefootmark{(d)} & $\alpha$\tablefootmark{(d)} & $ V $ & $ \Delta V $ & $ \HVa $ & Seeing\tablefootmark{(e)} & Weather\tablefootmark{(e)}\\
    YYYY-MM-DD & HH:MM:SS &  &  & $ \si{au} $ & $ \si{au} $ & $ \si{\degr} $ & $ \si{mag} $ & $ \si{mag} $ & $ \si{mag} $ & $ ^{\prime\prime} $ & \\
    \midrule
    2018-04-07 & 03:12:30 & CTIO & 7 & 3.8067 & 2.8184 & 2.7863 & 21.503 & 0.172 & 16.350 & 1.0--1.8 & 0--1 \\
    2018-04-07 & 12:40:57 & SSO & 13 & 3.8078 & 2.8183 & 2.6618 & 21.231 & 0.105 & 16.078 & 1.3--1.4 & 0 \\
    2018-04-07 & 21:43:13 & SAAO & 10 & 3.8088 & 2.8182 & 2.5429 & 20.916 & 0.089 & 15.762 & 1.4 & 0 \\
    \midrule
    2018-04-09 & 03:10:59 & CTIO & 9 & 3.8122 & 2.8182 & 2.155 & 20.794 & 0.092 & 15.638 & 1.5--1.8 & 0 \\
    2018-04-09 & 12:38:01 & SSO & 10 & 3.8132 & 2.8183 & 2.0303 & 20.633 & 0.099 & 15.477 & 1.2--1.4 & 0 \\
    \midrule
    2018-04-11 & 12:45:45 & SSO & 5 & 3.8187 & 2.8196 & 1.3952 & 20.490 & 0.086 & 15.330 & 1.3--1.5 & 0 \\
    2018-04-11 & 21:43:31 & SAAO & 11 & 3.8197 & 2.8200 & 1.277 & 20.432 & 0.103 & 15.271 & 1.8 & 0 \\
    \midrule
    2018-04-13 & 03:09:13 & CTIO & 4 & 3.8230 & 2.8216 & 0.8906 & 20.486 & 0.124 & 15.322 & 1.2--1.6 & 0--1 \\
    2018-04-13 & 21:44:11 & SAAO & 13 & 3.8250 & 2.8228 & 0.649 & 20.770 & 0.091 & 15.603 & 1.7 & 0 \\
    \bottomrule
  \end{tabular}
\tabnote{
\tablefootmark{(a)}{Average values of the UT date and time \textit{at the target} (i.e., corrected for the light travel time) of the images taken at the same night and the same observatory. The total observing time in each night was less than 1 hour, which is substantially smaller than the rotational and precessional periods of Toutatis ($ \sim 5 $--$ 7 \si{days} $).}
\\ \tablefootmark{(b)}{Telescope sites: Cerro Tololo Inter--American Observatory (CTIO), South African Astronomical Observatory (SAAO), and Siding Spring Observatory (SSO).}
\\ \tablefootmark{(c)}{Number of exposures.}
\\ \tablefootmark{(d)}{Averaged over the $ N $ ephemerides in a given night.}
\\ \tablefootmark{(e)}{Seeing and weather conditions from the official daily report (see Notebook 6 of Appendix \ref{app: online}). Weather codes 0 and 1 correspond to ``clear'' and ``thin clouds'', respectively.}
}
\end{table*}

The locations of Toutatis were specified using the Moving Object Detection Program (MODP) developed by KASI. MODP updates the World Coordinate System (WCS) information in the header of each image first and specifies the position of the asteroid in each image. For convenience, we analyzed images clipped to $ 2000 \times 2000 $ pixels ($ 800\arcsec \times 800\arcsec $) centered on the Toutatis locations calculated by the MODP. Cosmic-ray signals were removed with \texttt{astroscrappy}\footnote{\url{https://github.com/astropy/astroscrappy} version 1.0.5 using a separable median filter.}
which uses the \texttt{L. A. Cosmic} algorithm (\citealt{vanDokkum2001}). We coadded all images taken in the same night at the same observatory while shifting the individual images to increase the signal-to-noise (S/N) ratio. Eventually, two composite images per night per observatory were generated: one co-added image with sidereal offset and one by centering the individual images on Toutatis.

The signals from sources detected in each composite image are compared with those of field stars in the Pan-STARRS Data Release 1 (PS1, \citealt{PS1arxiv2016}). We first cone-searched PS1 objects within $ 0.06\degr $ centered at Toutatis and with more than 5 observations in the $g$ and $r$ bands.
We then used the criterion $ \texttt{iMeanPSFMag} - \texttt{iMeanKronMag} > 0.05 $ to reject non-stellar objects,\footnote{\url{ https://outerspace.stsci.edu/display/PANSTARRS/How+to+separate+stars+and+galaxies}} where \texttt{iMeanPSFMag} and \texttt{iMeanKronMag} are the mean i-band magnitudes obtained from using point spread function and Kron radius photometry, respectively. If saturated stars resulted in blooming and/or contaminated nearby pixels, all affected pixels were manually masked, and stars located closer than $ \sim 6 d $, where $ d $ is the full-width at half-maximum of the seeing disk, were rejected. We fixed $ d $ to be 4 pixels ($ \sim 1.6 \arcsec $). In addition, nearly colocated stars were identified and rejected by using the DAOGROUP algorithm (\citealt{StetsonDAOPHOT1987} implemented via \texttt{photutils} by \citealt{photutils_v0.4}) with a minimum separation distance of $ 7.5 d $.

Next, we used the transformation formula
\begin{equation} \label{eq: Lupton 2005}
  V = g - 0.59 (g - r) - 0.01
\end{equation}
from Lupton (2005)\footnote{\url{https://www.sdss.org/dr14/algorithms/sdssUBVRITransform/\#Lupton2005}} to obtain the Johnson--Cousins  $V$-band magnitudes of the stars, where $ g $ and $ r $ are the mean aperture photometry magnitudes (\texttt{MeanApMag}) of the $g$- and $r$-band, respectively. We fit a line to $ V $ and the instrumental magnitude from circular aperture photometry of the stars (aperture radius $ 1.5 d $, inner and outer sky radii $ 4 d $ and $ 6 d $, respectively) in the sidereally coadded images; only stars with $ 15 \le V \le 20$ are used for the analysis. The regression line provides the amount of atmospheric extinction in relative photometry, and we used it to obtain the V-band magnitude of Toutatis in the Toutatis-centered images via extrapolation. The minimum number of stars used in the regression (in the April 7 data from CTIO) was nine. Due to the extrapolation, the uncertainty in $ V $, $ \Delta V $ (obtained from the 1-$ \sigma $ confidence interval of the regression line), must be understood as a lower limit of the true uncertainty.

After data reduction, we compared the results with the raw images and found an artificial pattern contaminating one of the datasets (2018-04-11 SAAO). For this special case, we did manual photometry (selecting appropriate sky regions by visual inspection); all other data sets were analyzed as described above. We summarize the observational quantities in Table \ref{table:obslog} and plotted them as a function of time in Figure \ref{fig:halpha}.

\section{Derivation of the Geometric Albedo \label{sec: method}}

\subsection{Correction for Rotation and Distances \label{ss: shape}}

The observed $ V $ varies not only because of rotation and variation of the solar phase angle ($ \alpha $, the Sun--asteroid--observer angle) but also because of changes in the heliocentric and observer distances ($ r_\h $ and $ r_\g $, respectively). We first corrected the effects of the heliocentric and observer distances by deriving the reduced magnitude $ \HVa $, which is defined as a magnitude when the asteroid is viewed at unit heliocentric and observers distance (i.e., $ r_\h = r_\g = 1 \si{au} $) but at arbitrary phase angles. This parameter is given by
\begin{equation} \label{eq:reduced}
  \HVa = V (r_\h, r_\g, \alpha) - 5\lg \left ( \frac{r_\h r_\g}{1 \si{au} \cdot 1 \si{au}} \right ) ~.
\end{equation}
The IAU $ H $, $ G $ magnitude system \citep{BowellE+1989Asteroids2_IAU_HG} describes the $ \alpha $-dependence of $ \HVa $ using two parameters, the absolute magnitude $ \HV = \HV(\alpha = 0) $, and the slope parameter $ G $:
\begin{equation}\label{eq: IAU HG}
  \HVa = \HV + 2.5 \lg [(1 - G)\Phi_1(\alpha) + G \Phi_2(\alpha)]~,
\end{equation}
where
\begin{eqnarray}
    \Phi_1(\alpha) & = & \exp\left[-3.33 \tan^{0.63}(\alpha/2)\right] ~, \nonumber \\
    \Phi_2(\alpha) & = & \exp\left[-1.87 \tan^{1.22}(\alpha/2)\right] ~.
\end{eqnarray}

Our observations were made at very small $ \alpha $ values. For a spherical object, the fraction of dark area (outside the terminator) is $ f_\mathrm{dark} = (1 - \cos \alpha) / 2 $. Since our observations were made at $ \alpha < 3\degr $, $ f_\mathrm{dark} < 0.07 \si{\%} $, and a non-spherical shape, such as that of Toutatis, does not significantly increase this value. Therefore, we assume that the total projected area of the target viewed from a given direction will be the same as that of the sunlit surface.

Furthermore, the reflected intensity per projected area (which is closely related to the radiance factor; see Section \ref{ss: radf}) is nearly independent of the incidence and emission angles and depends only on the phase angle $ \alpha $, when the object is viewed near opposition. Thus, the flux reflected from a surface patch on Toutatis is proportional to its projected area, regardless of its orientation. This was first conjectured observationally by \cite{MarkovA+BarabashevNP1926_Lunar_RADF} from lunar observations as mentioned in the appendix of \cite{DollfusA+Bowell1971AnA_Apollo_Part1}. It was confirmed by \cite{LeeM+IshiguroM2018AnA} using a modern theoretical model: their Figure 4 shows that the radiance factor (defined in Section \ref{ss: radf}) is reduced only by $ \sim 1 \si{\%} $ or $ \sim 3 \si{\%} $ when the incidence angle is chagned from $ 0 \degr $ to $ 70 \degr $ or $ 90 \degr $ (the azimuthal angle and phase angle are fixed to 0). This means, if we correct the phase effect using Equation (\ref{eq: IAU HG}), the obtained $ \HV $ will be linearly related to the logarithm of the geometrical cross-sectional area (Equation \ref{eq: IF1} below).

Figure \ref{fig:halpha} shows $ \HVa $ and $ \HV $ as function of time, using $ G = 0.10 $ following \cite{1995Icar..117...71S}. The best rotational model, based on the high-resolution radar shape model of \citet{2004PDSS...16....8H}, is shown as red solid line, and other models within the 1-$ \sigma $ confidence interval are shown as faint black dotted lines. The process of selecting rotational models is described in Appendix \ref{app: rotation}. We confirmed that any $ G $ with $ 0.10 \le G \le 0.50 $ does not change our results. Our reduced magnitudes and the amplitude of the light curve agree with previously reported results \citep{1995Icar..117...71S}.

\subsection{Radiance Factor and Geometric Albedo \label{ss: radf}}

The radiance factor is given by $ I/F $ where $ I $ is the measured irradiance of the object, whereas $ F $ is that of a perfectly diffuse (Bond albedo of unity) Lambertian plate\footnote{Using a Lambertian \textit{plate} rather than a Lambertian \textit{sphere} has historical rather than mathematical or physical reasons \citep[see, e.g.,][]{RussellHR1916ApJ_Albedo}. If we used a Lambertian \textit{sphere}, the $ F $ value would be $ 1 / 2\pi $ rather than $ 1 / \pi $.} of the same projected area at the identical geometric configuration to that of the object, but oriented such that the incidence angle of sunlight is $ 0 $. Since the projected area is the same for both $ I $ and $ F $, the ratio $ I/F $ will be identical to the ratio of the bidirectional reflectance of the object (which is unknown) and the Lambertian plate (which is $ \cos(0) / \pi = 1 / \pi $). Thus, at a perfect opposition ($ \alpha = 0\degr $), $ I/F $ coincides with $ \pV $ by definition.

\begin{figure}[t!]
  \centering
  \includegraphics[trim=2mm 0mm 4mm 2mm, clip, width=1\linewidth]{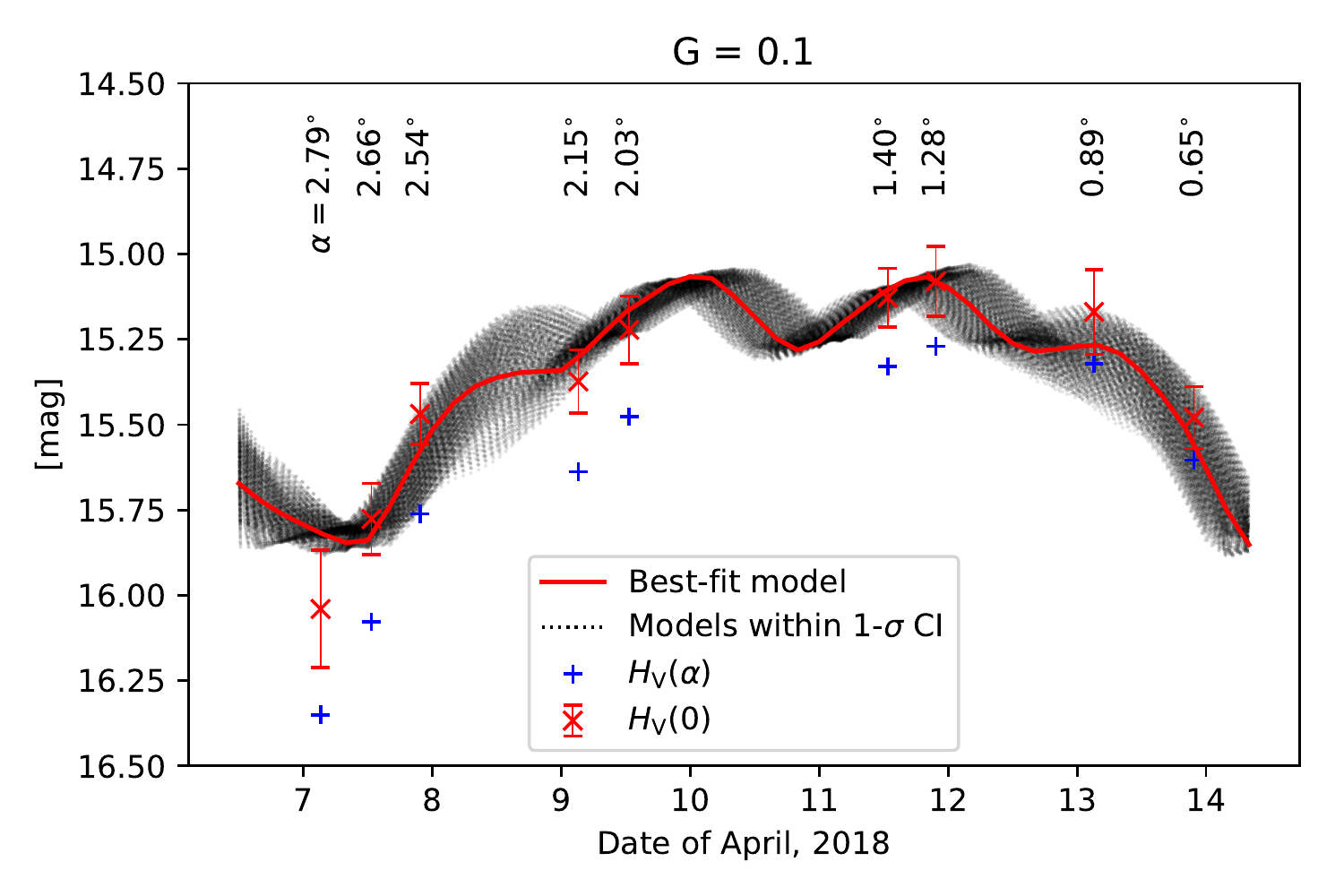}
  \caption{Reduced magnitude, $ \HVa $, and phase-corrected magnitude, $ \HV $, as function of time. The numbers above each marker indicate the phase angles $ \alpha $ at the observation epoch. The red solid line is the best-fit  geometrical cross-sectional area, $ -2.5 \lg (S_\proj (t)) $ (See Equation \ref{eq: IF1}), the faint black dotted lines indicate model curves within the 1-$ \sigma $ confidence interval (See Appendix \ref{app: rotation}). The short-time ($ \sim $1--2 days) bumpy features in the data are reproduced in the model, meaning that these features are attributed to the rotation of the asteroid.}
  \label{fig:halpha}
\end{figure}

As described in Section \ref{ss: shape}, the $ I/F $ value remains almost constant over the surface when the object is observed near opposition. Thus, the observed flux $ F_\mathrm{obs} $ is proportional to $ I/F (\alpha) \times S_\proj(t) $, where $ I/F(\alpha) \approx \mathrm{const} $ over the surface of Toutatis. The obtained reduced magnitude, $ \HVa$, can then be converted into a logarithm of $ I/F $:
\begin{equation} \label{eq: IF1}
 -2.5 \lg \hskip-2pt \left( \frac{I}{F} \right)\hskip-2pt
    = \hskip-2pt \HVa - V_\odot - 2.5 \lg \hskip-2pt \left( \frac{\pi}{S_\proj(t)} \right) + m_c ,
\end{equation}
where $ V_\odot = -26.762 $ is the V-band magnitude of the Sun at 1 au \citep{2015PASP..127..102M}, $ S_\proj (t) $ is the geometrical cross section in $ \si{m^2} $ as a function of time, and $ m_c = -5\lg (1 \si{au} / 1 \si{m}) = -55.87 $ is a constant to adjust the length unit.

\begin{figure}[t!]
  \centering
  \includegraphics[trim=4mm 2mm 4mm 4mm, clip, width=1\linewidth]{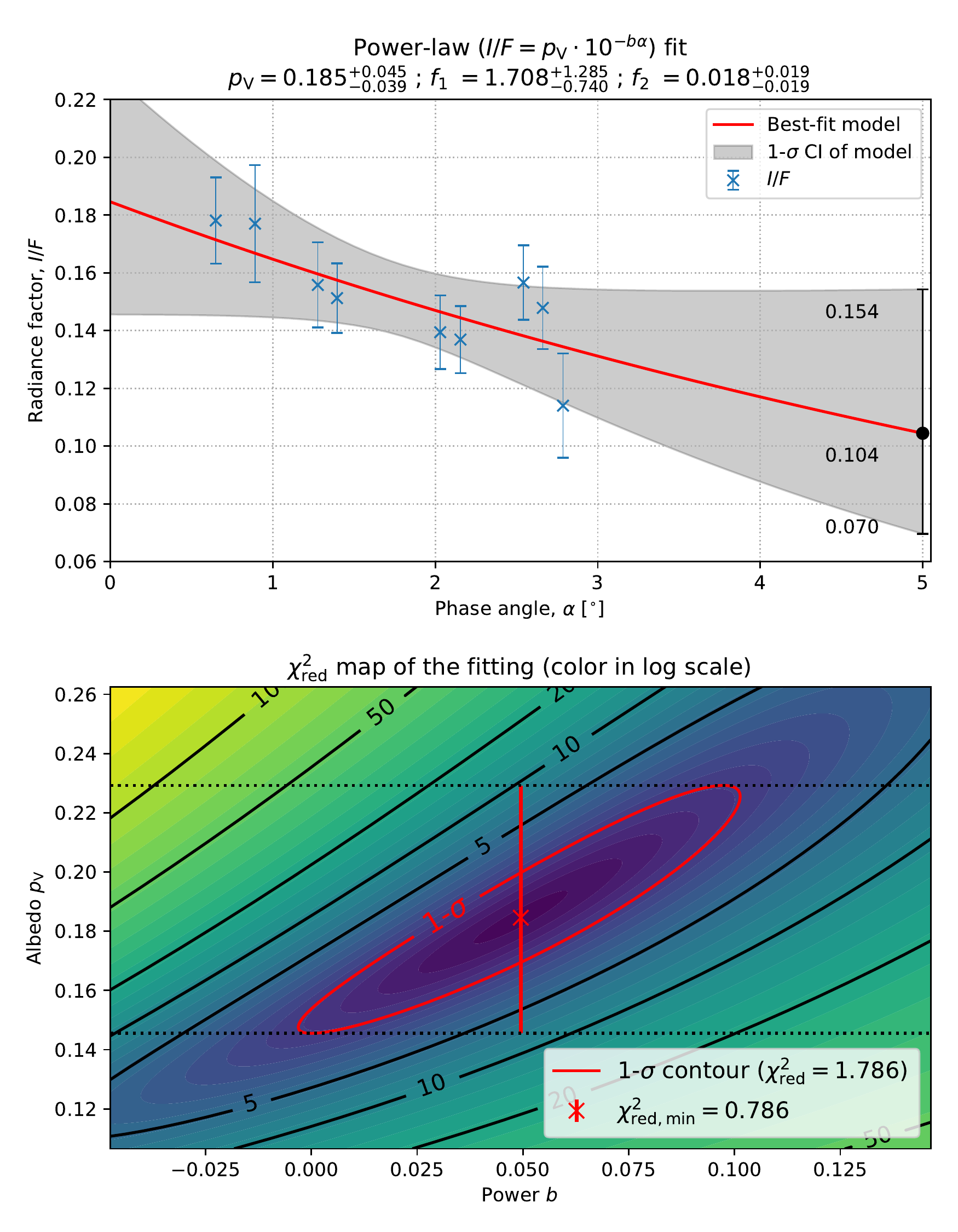}
  \caption{\emph{Top:} The radiance factor ($ I/F $) as a function of phase angle ($ \alpha $). The observed radiance factor values (blue cross markers) are fitted using a single power-law: $ I/F (\alpha) = \pV \times 10^{b \alpha} $. The best fit $ \pV $, $ f_1 $, and $ f_2 $ (see Equation \ref{eq: factors}) and their 1-$ \sigma $ uncertainties are shown above the figure. The red solid line is the best-fit model, the shaded area marks the 1-$ \sigma $ confidence interval of the model fit. The black marker shows $ A_5 = I/F (\alpha = 5\degr) $ and its 1-$ \sigma $ confidence interval, indicated by the numbers next to the marker. \emph{Bottom:} The corresponding $ \chi^2_\mathrm{red} $ statistic in the parameter space. The 1-$ \sigma $ confidence contour of $ \chi^2_\mathrm{red} $ and that of $ \pV $ are shown. The colormap is in log-scale and contours for a few selected $ \chi^2_\mathrm{red} $ values are shown to guide the eyes.}
  \label{fig:figradf}
\end{figure}

The calculated $ I/F $ values are plotted against the phase angle $ \alpha $ in the upper panel of Figure \ref{fig:figradf}. We used a single power-law ($ I/F (\alpha) = \pV \times 10^{b \alpha} $) to fit the points approximately. From this plot, we considered four parameters of interest: the geometric albedo ($ \pV $), the albedo at phase angle of $ 5\degr $ ($ A_5 $), and two factors
\begin{equation}
  f_1 = \frac{I(0.3\degr)}{I(5\degr)}
  \quad;\quad
  f_2 = \frac{I/F(0.7\degr) - I/F(2.5 \degr)}{2.5 \degr - 0.7 \degr} ~,
\end{equation}
which were used to check the validity of our results by comparing with previous works (explained below).

We used a brute-force grid search algorithm (using a fixed grid) to find the best-fit power-law parameters $ \pV $ and $ b $. The best-fit parameters were first found by simple $\chi^2$ minimization where the reduced $\chi^2$ is defined as
\begin{equation}
  \chi^2_\mathrm{red} :=
    \frac{1}{N_\mathrm{obs} - 2} \sum_{i=1}^{N_\mathrm{obs}}
      \left ( \frac{I/F(\alpha_i) - \pV \times 10^{b \alpha_i}}{\sigma_i} \right )^2 .
\end{equation}
Here, $ N_\mathrm{obs} = 9 $ is the total number of data points, $ N_\mathrm{obs} - 2 = 7 $ is the number of degrees of freedom, and $ \sigma_i $ is the uncertainty in $ I/F $ for the $ i $-th observation. The $ \chi^2_\mathrm{red} $ contours in the 2-D parameter space (bottom panel of Figure \ref{fig:figradf}) are carefully investigated to assure that we searched a large enough parameter space with a sufficiently fine grid. The uncertainties of $ f_1 $, $ f_2 $, and $ A_5 $ are estimated based on the minimum and maximum values of each of them for the models within the 1-$ \sigma $ confidence interval (the gray region in the top panel of Figure \ref{fig:figradf}).

From the fit, the amplitude parameter is the geometric albedo by definition ($ \pV \equiv I/F(0) $). We obtained the geometric albedo in the V-band $ \pV = \pVval $. This is a typical albedo value for an S-type asteroid.

The albedo at $ \alpha = 5\degr $, $ A_5 $, is important since this is used in experimental studies to compare with celestial bodies (see Section \ref{ss: polarimetry}). It is defined as the ratio of the measured flux of the obeject of interest at $ \alpha = 5\degr $ to that of a perfectly diffuse Lambertian plate at the same geometrical configuration with the same projected area.\footnote{The angle of $ \alpha = 5\degr $ is chosen by experimentalists, since the albedo measurements were found to give the most reproducible results with $ \alpha = 5\degr $ \citep[see][and references therein]{GeakeJE+DollfusA1986MNRAS}. A smoked MgO screen illuminated at an incidence angle of $ 0 \degr $ is used as an analog of Lambertian plate. The $ A_5 $ is the ratio of the flux of the target of interest and the MgO screen, while both fluxes are measured at $ \alpha = 5 \degr $ with zero incidence angle.} Thus,
\begin{equation}\label{eq: A_5}
  A_5 = \frac{I}{F} (\alpha = 5\degr) = \Aval ~,
\end{equation}
as the black marker shown in Figure \ref{fig:figradf}.
The two factors, $ f_1 $ and $ f_2 $ are
\begin{equation} \label{eq: factors}
  f_1 = \faval
  \quad;\quad
  f_2 = \fbval \si{/\degr} ~.
\end{equation}
$ f_1 $ was investigated in \cite{BelskayaIN+ShevchenkoVG2000Icar_OE} and found to be $ f_1 = 1.34 \pm 0.05 $ for C-type and $ f_1 = 1.44 \pm 0.04 $ for S-type asteroids. From the V-filter result in Figure 7 of \cite{TatsumiE+2018Icar_Itokawa_OE}, $ f_1 \sim 1.3 $ for the asteroid (25143) Itokawa, an S-type asteroid.
$ f_2 $ was determined for asteroid Itokawa in \cite{LeeM+IshiguroM2018AnA}. For Itokawa, they found $ f_2 \sim 0.01 \pm 0.01 $ for $ \alpha > 1.4\degr $ and $ f_2 \sim 0.04 \pm 0.01 $ for $ \alpha < 1.4\degr $ (note that the sign convention is opposite from the one for their $ s_\mathrm{OE} $). We employed $ 0.7 \degr $ and $ 2.5 \degr $ because the phase angle difference is larger than $ 0.3 \degr $ as \cite{LeeM+IshiguroM2018AnA} and this is the interpolated region. Both $ f_1 $ and $ f_2 $ coincide with the expected values from previous works, but it is difficult to arrive at further conclusions due to the uncertainties.

Our parameter values are summarized in Table~\ref{table:results}. In the last column of the table, we summarized the values of the same parameters but derived by excluding one of the data points (and updating the number of degrees of freedom accordingly), 2018-04-07 CTIO with $ \alpha = 2.79 \degr $, since this is the observation with the largest deviation from the shape model prediction (Figure \ref{fig:halpha}). Comparing the two results, it can be seen that the derived parameters are robust against the exclusion of the deviant data point (parameters changed much less than the 1-$ \sigma $ confidence range).

\begin{table}[t!]
\centering
\caption{Parameter values derived using either all observational data or after excluding the data of 2018-04-07 CTIO ($ \alpha = 2.79 \degr $)}
\label{table:results}
\setlength{\tabcolsep}{12pt}
  \begin{tabular}{ccc}
    \toprule
    Parameters & All data & Exclude $ \alpha = 2.79\degr $\\
    \midrule
    $ \pV $ & \pVval & $ 0.179^{+0.042}_{-0.037} $ \\
    \addlinespace
    $ A_5 $ & \Aval  & $ 0.116^{+0.055}_{-0.039} $\\
    \addlinespace
    $ f_1 $ & \faval & $ 1.509^{+1.118}_{-0.649} $ \\
    \addlinespace
    $ f_2 \si{[/\degr]}$ & \fbval & $ 0.014^{+0.019}_{-0.019} $ \\
    \bottomrule
  \end{tabular}
\end{table}

\section{Discussion}

The geometric albedos of asteroids have been derived with several techniques,  including (1) infrared observations of  thermal radiation, (2) time-delay and Doppler measurements of radar signals emitted from ground-based stations, (3) measurements of the time duration during occultation of background stars, (4) direct imaging with adaptive optics techniques, (5) \textit{in situ} measurements of the bidirectional reflectance using spacecraft cameras, and (6) an empirical method using polarimetric slopes. Since  techniques (1)--(4) are contrived to derive the sizes (and shapes) rather than the albedo values of asteroids, it is essential to combine the visible magnitudes to derive the albedos using these four techniques. It is, however, true that optical observations have not been conducted strenuously despite recent comprehensive surveys, especially  by means of (1) and (2). In general, the visible magnitudes have been obtained using uncalibrated observation data on the presumption that asteroids have a phase function (typically $ G = 0.15 $ for an IAU $ H $, $ G $ magnitude system). In fact, it is quite difficult to make optical observations of asteroids at the opposition point (or at least within a hypothetical circle having the equal apparent diameter of the Sun) because the orbital planes of almost all asteroids do not coincide with Earth's orbital plane (i.e., $ i \neq 0 $). Fast time variations of the cross-sectional areas of asteroids also make it difficult to derive the albedo values even when the phase function and dimension of the asteroids are known. This effect is especially significant for elongated asteroids. In contrast, direct observations of asteroidal surfaces via space missions provide more reliable information about the albedos, eliminating the above uncertainties regarding the cross-sectional areas. However, note that the calibration processes of onboard data have some uncertainties, such as the standardization of magnitude systems and unavailability of flat-fielding data in space. Some \textit{in situ} data were acquired at large phase angles (e.g., (433) Eros by the NEAR-Shoemaker mission). Thus, it is not straightforward to derive the accurate geometric albedo values.

\subsection{Sources of Uncertainties \label{ss: uncertainties}}

The parameters we obatained in this study, especially the albedos, may have suffered from errors that are difficult to quantify correctly. For one of our observations (2018-04-07 CTIO with $ \alpha = 2.79\degr $), the magnitude is $ \sim 1 $-$ \sigma $ fainter than the model prediction (Figure \ref{fig:halpha}). Although a 1-$ \sigma $ of deviation is not necessarily significant, there is a possibility that the value actually was affected by an unknown systematic shift, such as those caused by imperfect phase function modeling (Equation \ref{eq: IAU HG}) and weather effects (Table \ref{table:obslog}). Although it was almost invisible, some artifact as in 2018-04-11 SAAO (see the last part of Section \ref{sec: observation}) may have affected this image in a non-trivial way. This data, however, does not seem to critically affect the quantities derived in this work, as shown in Table \ref{table:results}.

In addition, as the radiance factor calculation requires the absolute physical value of the projected area (Equation \ref{eq: IF1}), the uncertainty in the physical size affects the $ I/F $ value. From Equation (\ref{eq: IF1}), $ I/F \propto 1 / S_\proj $ ($ \sim D^{-2} $ to first order), if all other quantities including the observed values are kept constant. The radar models we used give the maximum sizes along each principal axes as $ (4.60,\, 2.28,\, 1.92) \pm 0.10 \si{km} $ \citep{2003Icar..161..346H}, which is slightly smaller but within the uncertainty ranges obtained by Chang'e-2, $ (4.75 \pm 0.48,\, 1.95 \pm 0.19) \si{km} $ \citep{2013NatSR...3E3411H}. Ignoring the uncertainty ranges, $ S_\proj $ may have been underestimated systematically by $ \sim 5 \si{\%} $ in our estimation. Hence, our albedo estimations, $ \pV \equiv I/F(\alpha = 0) $ and $ A_5 \equiv I/F(\alpha = 5\degr) $, could have been overestimated by $ 0.009 $ and $ 0.005 $, respectively. These uncertainties are negligible compared to the statistical uncertainties (only $ \sim 10 \mathrm{-} 25 \si{\%}$ of the error-bars; see Table \ref{table:results}).

\subsection{Comparison with Flyby Observations \label{ss: chang'e-2}}

The Chinese Chang'e-2 spacecraft succesfully obtained optical images of Toutatis during its December 2012 flyby at a distance of $ \sim 1.3 \si{km} $, and the hemispherical albedos were derived \citep{ZhaoDF+2016AcASn_Change_hemisphr_albedo}. These results allow testing of the albedo estimates gained from different approaches including this work. It is, however, difficult to directly compare their results to ours since they estimated the hemispherical albedo (assumed to be equal to the Bond albedo, $ A_\mathrm{B} $), different from bidirectional albedos derived in this work (e.g., $ \pV $ and $ A_5 $).

Using the IAU H, G system \citep{BowellE+1989Asteroids2_IAU_HG}, the Bond albedo can be obtained from $ \pV $ using a numerical approximation of the phase integral:
\begin{equation}\label{eq: Bond_Albedo}
  A_\mathrm{B, V} = (0.286 + 0.656 G) \pV ~,
\end{equation}
where $ A_\mathrm{B, V} $ is the Bond albedo in V-band.\footnote{Conventionally, $ A_\mathrm{B, V} = (0.290 + 0.684 G) \pV $ suggested by \cite{BowellE+1989Asteroids2_IAU_HG} is widely being used. However, \cite{MyhrvoldN2016PASP} found that the best fitting function is $ A_\mathrm{B, V} = (0.286 + 0.656 G) \pV $. We confirmed the latter result and used it here.}
Using the uncertainty in $ \pV $ and $ G = 0.0 \mathrm{-} 0.2 $, we obtain extrema of $ A_\mathrm{B, V} = 0.04 \mathrm{-} 0.10 $. This is smaller than the values derived in \cite{ZhaoDF+2016AcASn_Change_hemisphr_albedo}: 0.2083, 0.1269, and 0.1346 in the R, G, and B bands of Chang'e-2, respectively. Using Equation (\ref{eq: Bond_Albedo}), these can be transformed into the geometric albedos of $ 0.50 \mathrm{-} 0.73 $, $ 0.30 \mathrm{-} 0.44$, and $ 0.32 \mathrm{-} 0.47 $ assuming $ G = 0.0 \mathrm{-} 0.2 $ and identical phase functions, respectively.

The discrepancy may have been caused by the assumptions used to compare the two physically different albedos (bidirectional and Bond). To measure the Bond albedo directly, one needs to observe the object at all the possible $ 4\pi \si{sr} $ directions, which is impractical. Thus, we assumed a phase function (Equation \ref{eq: Bond_Albedo}) to convert the $ \pV $ to $ A_\mathrm{B, V} $, while \cite{ZhaoDF+2016AcASn_Change_hemisphr_albedo} assumed illumination conditions (incidence angles) to convert the radiance ($ \mathrm{W \, m^{-2} \,  sr^{-1} \, \mu m^{-1}} $) into the albedo, for example. The phase angle coverages of the observations in both works are also different.

Therefore, this discrepancy is a result of imperfect assumptions used to indirectly infer the Bond albedo in the two works, while these systematic uncertainties are not taken into account in the error bars. We emphasize that our $ \pV $ and $ A_5 $ should be accurate within the derived uncertainty range, because these were obtained from observations made at $ \alpha = 0.6 \degr \mathrm{-} 2.8 \degr $, close to where these quantities are defined.

\subsection{Comparison with Thermal Infrared Studies \label{ss: neowise}}

The sizes and albedos provided by infrared catalogs such as IRAS \citep{2002AJ....123.1056T}, AKARI \citep{2011PASJ...63.1117U}, and NEOWISE \citep{2017AJ....154..168M} provide an invaluable overview on asteroids of different taxonomic types distributed across the solar system; nonetheless, it is also important to note that some asteroids with a small number of detections and elongated shapes have large uncertainties due to rotation.

\citet{2017AJ....154..168M}\footnote{The machine-readable table is available at \url{http://iopscience.iop.org/1538-3881/154/4/168/suppdata/ajaa89ect1_mrt.txt}} obtained $ \lg \pV = -0.392 \pm 0.166 $ with a diameter $ D = 1.788 \pm 0.376 \si{km} $ for Toutatis from  6 NEOWISE infrared observations in both W1 and W2 bands at phase angle $ \alpha \sim 48\degr $, in combination with $ H = 15.30 $, $ G = 0.10 $, and a beaming parameter $ \eta = 1.007 \pm 0.223 $. The $ \pV $ value can be transformed to $ \pV = 0.406^{+0.189}_{-0.129} $, if we assume the posterior distribution of their $ \lg \pV $ estimation is truly, or at least similar to, Gaussian. It should be noted that the observations were made at a single phase angle, so the Near Earth Asteroid Thermal Model (NEATM) used in \citet{2017AJ....154..168M} may not give an accurate estimation, but other advanced modelings may be more adequate, such as the advanced thermophysical model (ATPM; \citealt{RozitisB+GreenSF2011MNRAS_ATPM}, which is used for space mission related targets, e.g., \citealt{YuLL+2014MNRAS_439_3357, YuLL+2015MNRAS_452_368}).

Meanwhile, \citet{2014PASJ...66...56U} noted a 22\% deviation for the 1-$ \sigma $ confidence range among three radiometric survey catalogs by IRAS, AKARI, and WISE \citep{2002AJ....123.1056T, 2011PASJ...63.1117U, 2011ApJ...731...53M}. The deviation is caused not only by the different types of thermal models but also by the observed rotational phases of the asteroids. It is worth noting that the size and albedo determination can be more uncertain for individual targets than the estimated 22\% from the \textit{average} trend.

Note also the currently ongoing debate about the accuracy of the NEOWISE results \cite[see, e.g., ][]{MyhrvoldN2018Icar314, WrightEL+2018arXiv, MyhrvoldN2018arXiv}. According to \cite{WrightEL+2018arXiv}, however, the results we used for Toutatis \citep{2017AJ....154..168M} are not affected by any problem. Meanwhile, \cite{MyhrvoldN2018Icar303} discusses a new approach to the NEATM by introducing a new parametrization and a physically valid yet simple generalization for the case that the reflected sunlight is not negligible in the thermal spectra. This new approach may improve the NEOWISE results with respect to accuracy and consistency.

In case of Toutatis, one needs to pay close attention to the rotational phase because of the very elongated shape of the asteroid. In fact, the size given in the NEOWISE catalog is only 73 \% of the mean diameter derived from radar observations (i.e., $ D = 2.45 \si{km} $). Using the relationship $ D = 1329 \si{[km]} \pV^{-0.5} 10^{-\HV / 5} $ \citep[derived in, e.g., the appendix of ][]{PravecP+HarrisAW2007Icar}, the $ \pV $ value appears to be overestimated by a factor of $ \sim 1 / 0.73^2 = 1.88 $. Despite this not being an accurate statement since the change in $ \pV $ changes the shape of the thermal spectrum of the asteroid, we may crudely estimate the updated albedo to be $ \pV \sim 0.406 / 1.88 = 0.22 $. This coincides with the albedo expected for an S-type asteroid.

\subsection{Comparison with Polarimetric Studies \label{ss: polarimetry}}

\citet{1995Icar..113..200L} studied the polarimetric properties of Toutatis and derived $\pV = 0.13$ using the empirical polarimetric slope--albedo law stated in \citet{1977LPSC....8.1091Z}. Subsequently, \citet{1997Icar..127..452M} derived the same value, $ \pV = 0.13 $, using the polarimetric slope--albedo law in \citet{1979aste.book..170D}. The empirical relationship, first noted in \citet{widorn1967}, is
\begin{equation} \label{eq:zellner1979}
  \lg \pV = C_1 \lg h + C_2 ~,
\end{equation}
where $ h $ denotes the polarimetric slope around the inversion angle in $ \si{\% / \degr} $ and where $C_1$ and $C_2$ are constants. Note that the albedo in \citet{1977LPSC....8.1091Z} and \citet{1979aste.book..170D} is different from the one that we currently apply. Thus, these authors derived a set of constants $ (C_1,\, C_2) = (-0.93,\, -1.78) $ based on the laboratory polarimetric measurements, where the measurements defined the albedo at the normal ($ 0\degr $) incidence and $ 5\degr $ angle of emergence (i.e., $ \alpha = 5\degr $, not $ 0\degr $). It is challenging to measure the geometric albedo, which is defined at $ \alpha = 0\degr $, because the light source is in the line of sight of the detector, thus hiding the sample. Although we acknowledge these early efforts, the polarimetric albedo in \citet{1995Icar..113..200L} and \citet{1997Icar..127..452M} needs to be updated to match the definition of the geometric albedo.

Multiple studies made efforts to derive the set of constants for obtaining the geometric albedo from the polarimetric slope \citep{2012ApJ...749..104M, 2015MNRAS.451.3473C, 2018SoSyR..52...98L}, involving better ways to calibrate the data from different sources. We estimated the geometric albedos using the polarimetric slope--albedo law using these sets of parameters, and the results are summarized in Table \ref{table:polalbedo}. To first order, the Gaussian 1-$ \sigma $ confidence range of $ \pV $ is estimated from Equation (\ref{eq:zellner1979}) and
\begin{equation}\label{key}
  \Delta \pV
    \sim \sqrt{ \left ( \pdv{\pV}{C_1} \right )^2 \hskip-5pt \Delta C_1^2
    + \left ( \pdv{\pV}{C_2} \right )^2 \hskip-5pt \Delta C_2^2
    + \left ( \pdv{\pV}{h} \right )^2 \hskip-5pt \Delta h^2} ~,
\end{equation}
where $ \Delta X $ is the Gaussian 1-$ \sigma $ confidence range for the parameter $ X $. We summarize the geometric albedo derived using different observations and different sets of parameters in Table \ref{table:polalbedo}. These polarimetric results are consistent with the geometric albedo that we derived through the direct photometric observation at opposition.

\begin{table*}[t]
\centering
\caption{Geometric albedo ($ \pV $) values obtained from polarimetry }
\label{table:polalbedo}
\setlength{\tabcolsep}{12pt}
  \begin{tabular}{l c c c c}
    \toprule
    Reference & $ C_1 $ & $ C_2 $ & \cite{1995Icar..113..200L} & \cite{1997Icar..127..452M} \\
    \midrule
    \citealt{2012ApJ...749..104M} & $ -1.207 \pm 0.067 $ & $ -1.892 \pm 0.141 $ & $ 0.182 \pm 0.066 $ & $ 0.186 \pm 0.073 $ \\
    \citealt{2015MNRAS.451.3473C} & $ -1.111 \pm 0.031 $ & $ -1.781 \pm 0.025 $ & $ 0.190 \pm 0.022 $ & $ 0.194 \pm 0.034 $ \\
    \citealt{2018SoSyR..52...98L} & $ -0.989 \pm 0.047 $ & $ -1.719 \pm 0.040 $ & $ 0.168 \pm 0.025 $ & $ 0.171\pm 0.033 $ \\
    \bottomrule
  \end{tabular}
  \tabnote{$ C_1 $ and $ C_2 $ are constants in the polarimetric slope--albedo law given by $\lg \pV = C_1 \lg h + C_2$. These parameters are determined by the workscited in the left column. We used the slope parameters given in \cite{1995Icar..113..200L} and \cite{1997Icar..127..452M}.}
\end{table*}

\begin{figure}[t!]
  \centering
  \includegraphics[trim=5mm 3mm 5mm 4mm, clip, width=1\linewidth]{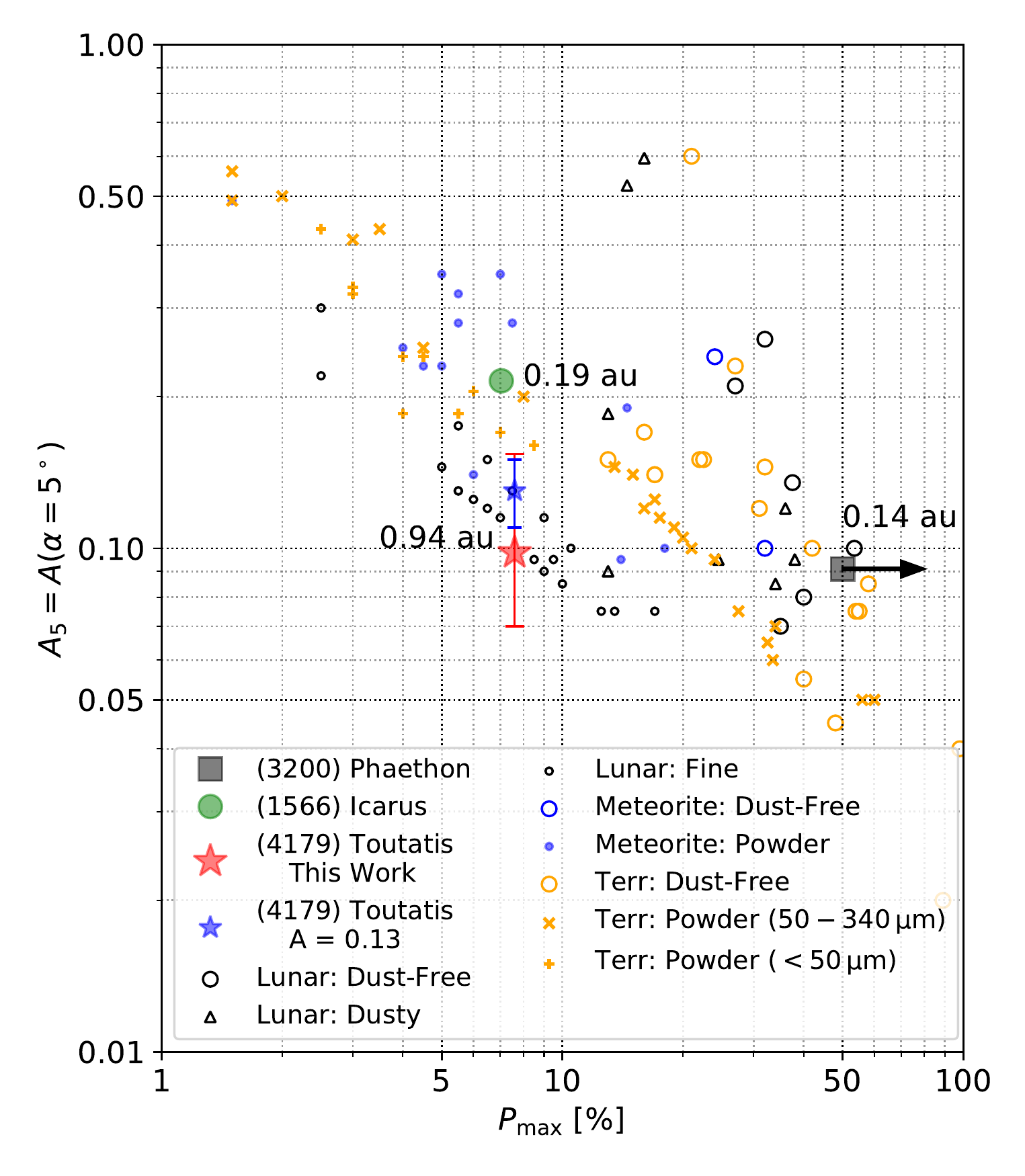}
  \caption{Albedo at $ \alpha = 5\degr $ vs. $P_\mathrm{max}$ plot for lunar and terrestrial samples in \cite{GeakeJE+DollfusA1986MNRAS} along with those of three asteroids whose $P_\mathrm{max}$ values are well studied. The data for (3200) Phaethon and (1566) Icarus are from \cite{IshiguroM+2017AJ_1566_Icarus} and \cite{ItoT+2018NatCo_3200_Phaethon_Pmax}, respectively. The word ``Terr'' in the legend means ``terrestrial'' materials. The numbers near the three asteroidal objects denote the perihelion distances of each asteroids in au.}
  \label{fig:grainsize}
\end{figure}

Finally, we derived the grain size on the surface. The empirical relationships regarding the grain sizes \citep{GeakeJE+DollfusA1986MNRAS} include $ A_5 $, not $ \pV $. In Section \ref{ss: radf}, we derived $ A_5 = \Aval $ in the V-band. From previous polarimetric data, it has been found that $ A_5 = 0.13 \pm 0.02 $ via the slope--albedo law \citep{1995Icar..113..200L, 1997Icar..127..452M}. By comparing with terrestrial rock powders (Figure \ref{fig:grainsize}), the grain size is estimated as $ d \lesssim 50 \si{\mu m} $.

The grain size is smaller than that of (3200) Phaethon ($ d \gtrsim  360 \si{\mu m} $,  \citealt{ItoT+2018NatCo_3200_Phaethon_Pmax}) and (1566) Icarus ($ d \sim 100 - 130 \si{\mu m} $, \citealt{IshiguroM+2017AJ_1566_Icarus}) that were calculated by using the same relationship. In addition, we compare the $ P_\mathrm{max} $--$ A_5 $ relationship of these three asteroids and laboratory sample data (Figure \ref{fig:grainsize}). It is likely that (3200) Phaethon and (1566) Icarus are covered with grains larger than that of Toutatis, although other factors such as the porosity of grains can affect the size estimate. These results may suggest that the near-Sun environment at less than approximately $ 0.2 \si{au} $ affects the paucity of small grains on the surface.

Note that the observations or calculations of $ A_5 $ and $ P_\mathrm{max} $ for these asteroids were sometimes made at wavelengths different from those used in the laboratory experiments in \cite{GeakeJE+DollfusA1986MNRAS}. Nevertheless, the deviations that can be expected are within the measurement uncertainties.

\section{Conclusions}

In this work, we analyzed imaging data of the $ \mathrm{S_k} $-type asteroid (4179) Toutatis at near-opposition configuration obtained by KMTNet. We obtained a robust geometric albedo value of $ \pV = \pVval $ from the observed radiance factors and the shape model. Our value is typical for an S-complex asteroid, and thus shows the future potential of using KMTNet for the opposition monitoring of asteroids. Other measures of albedo using the radiance factors ($ f_1 $ and $ f_2 $), which were introduced in previous works, agree with our results within the uncertainties albeit it is substantial. Thermal modeling and polarimetry give consistent results if we use appropriate diameter values or empirical relationships, respectively. The albedo at $ \alpha = 5 \degr $ is also determined from the radiance factor and is consistent with Toutatis being covered with particles smaller than the asteroids in the near-Sun environment found from empirical relationships involving the grain size.

%%% ACKNOWLEDGMENTS (IF ANY) %%%%%%%%%%%%%%%%%%%%%%%%%%%%%%%%%%%%%%%%

\acknowledgments

We thank the two anonymous reviewers for their valuable comments which led to important improvements of the manuscript.
This research has made use of the KMTNet system operated by the Korea Astronomy and Space Science Institute (KASI) and the data were obtained at three host sites of CTIO in Chile, SAAO in South Africa, and SSO in Australia. The observations around the opposition were made using telescopes in the frame of contract research between Seoul National University (SNU) and the Korea Astronomy and Space Science Institute (KASI) for the DEEP--South project. This research was supported by the Korea Astronomy and Space Science Institute under the R\&D program supervised by the Ministry of Science, ICT and Future Planning.

%%% APPENDICES (IF ANY) %%%%%%%%%%%%%%%%%%%%%%%%%%%%%%%%%%%%%%%%%%%%%
\appendix
\section{Rotational Model\label{app: rotation}}

We calculated the projected geometrical cross-sectional area, $ S_\proj (t; \varphi_1, \varphi_2) $ of the asteroid as seen from Earth at the epoch of the observations ($ t $), assuming arbitrary rotational and precessional offsets, $ \varphi_1 $ and $ \varphi_2 $. In the calculation, we employed the high-resolution shape model\footnote{\url{https://sbn.psi.edu/pds/resource/rshape.html}} \citep{2003Icar..161..346H, 2004PDSS...16....8H}. We considered the long-axis mode rotation and precession along $ (\lambda,\, \beta) = (180 \degr,\, -52 \degr) $ in the ecliptic coordinate \citep{1995Sci...270...84H}. Although these rotation and precession periods have been reported to be $ P_1 = 5.38 \si{days} $ and $ P_2 = 7.40 \si{days} $, respectively \citep{ZhaoY+2015MNRAS_Toutatis_spin_Change2}, it is necessary to perform fine tuning by adding the offsets $ \varphi_1 $ and $ \varphi_2 $ to fit the observed magnitudes to the modeled ones because the periods in the model are not determined well enough to be extrapolated to the epoch of our observations (uncertainty in periods multiplied by the time difference is much larger than the periods themselves). We thus defined the rotational angle along the long-axis and the precession axis ($ \theta_1 $ and $ \theta_2 $, respectively) as follows:
\begin{equation}\label{eq:theta}
  \theta_j (t,\, \varphi_j) = \frac{2 \pi (t - t_0)}{P_j} + \varphi_j ~,
\end{equation}
for $ j = 1, \, 2 $, where $ t_0 $ denotes the time of the reference epoch.

\begin{figure}[t]
  \centering
  \includegraphics[trim=3mm 4mm 9mm 5mm, clip, width=1\linewidth]{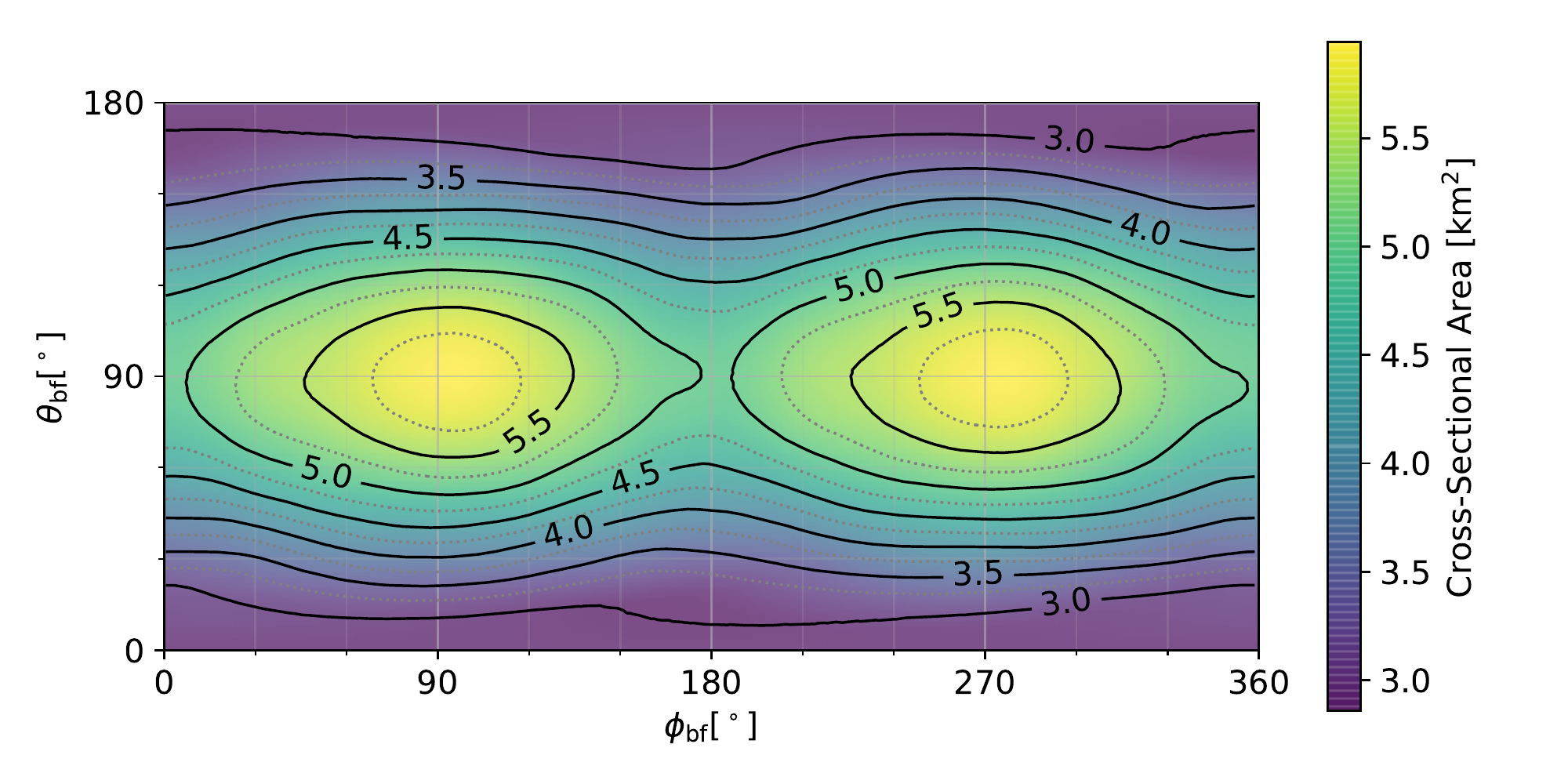}
  \caption{The cross-sectional area seen by an observer at infinite distance towards the direction of $ (\theta_\mathrm{bf}, \, \phi_\mathrm{bf}) $, described both in color and contour. $ \theta_\mathrm{bf} = 0 $ is the direction to the positive $ Z $-axis specified by the shape model. Cross-sections are in units of km$^2$.}
  \label{fig:figcrosssectionalarea}
\end{figure}

We used IDL \texttt{POLYSHADE} to calculate the total projected area $ S_\proj (t; \varphi_1, \varphi_2) $ for a given pair of $ (\varphi_1,\, \varphi_2) $ with grid size of $ \Delta \varphi_1 = 5 \degr $ and $ \Delta \varphi_2 = 2 \degr $ and time step size of $ 1/3 $ days. The projected area as a function of the direction to the observer is shown in Figure \ref{fig:figcrosssectionalarea}. The color and contour show the projected area viewed by an observer at an infinite distance to the direction specified the body-fixed frame, $ (\theta_\mathrm{bf}, \, \phi_\mathrm{bf}) $. As can be seen, an uncertainty of $ \sim 5 \degr $ in the viewing direction causes a fractional uncertainty (normally $ \ll 10 \si{\%} $) which is much smaller than the photometric uncertainty of $ \HVa $ in Figure \ref{fig:halpha} and Table \ref{table:obslog}. Thus, the choices for the grid size of the offsets ($ \Delta \varphi_1$ and $ \Delta \varphi_2 $) are reasonable.

Each of the 12,960 models for $ (\varphi_1, \, \varphi_2) $ pair is then spline interpolated when necessary. As mentioned in Section \ref{ss: shape}, the total reflected flux is proportional to the projected area when $ \alpha $ is small. Hence, for each rotational model, we fit $ a $ such that $ I_0(t) = a S_\proj(t) $, where $ I_0 (t) = 10^{-\HV(t) / 2.5} $ and its uncertainty is $ \Delta I_0(t) = \ln(10) I_0 (t) \Delta V(t) / 2.5 $ (using Equation (\ref{eq: IAU HG}) with fixed $ G $). Here, $ \HV(t) $ is the phase-corrected absolute magnitude, $ \HV $, calculated using Equation (\ref{eq: IAU HG}) at epoch $ t $. We used intensity units rather than magnitudes since  uncertainties in magnitude are not Gaussian, precluding the use of $\chi^2$ minimization. The reduced $ \chi^2 $ statistic is calculated for each case by using
\begin{equation}\label{eq: chisqred}
  \chi^2_\mathrm{red} (\varphi_1, \varphi_2) =
    \frac{1}{N \hskip-2pt - \hskip-2pt 1} \sum_{i=1}^{N}
    \left ( \frac{I_0(t_i) - a S_\proj(t_i; \varphi_1, \varphi_2)}{\Delta I_0(t_i)} \right )^2 ,
\end{equation}
where the denominator $ N - 1 $ is the number of degrees of freedom, $ N = 9 $ is the number of data points, and $ t_i $ is the $ i $-th observational time (Table \ref{table:obslog}). By varying $ \varphi_1 $ and $ \varphi_2 $, we find an $ a_0 $ for which $ a_0 = \arg \min_a \chi^2_\mathrm{red} (\varphi_1, \varphi_2) $. The minimum $ \chi^2_\mathrm{red} $ is then saved for each model.

The ``good'' models are selected such that the reduced $ \chi^2 $-statistics is smaller than $ 1 + \chi^2_\mathrm{red,\, min} $, where $ \chi^2_\mathrm{red,\, min} $ is the minimum of such reduced $ \chi^2 $ statistics of all the models. Because we have only one free parameter ($ a $), this gives the models with 1-$ \sigma $ confidence interval. The models are then converted into magnitude system using $ -2.5 \lg (a I_0) $. The minimum $ \chi^2 $ model is the red solid line and other ``good'' models are black faint dotted lines in Figure \ref{fig:halpha}. We confirmed that changes of $ G $ in the range $ G \in [0.10,\, 0.50] $ do not significantly distort the results, except for a nearly parallel shift caused by the nearly constant change (since all $ \alpha $'s are similar) in all $ \HV $ values (Equation \ref{eq: IAU HG}).

\section{On-Line Material \label{app: online}}

This work was carried using mostly open-source software, and we here disclose the codes we used to obtain our results. The original codes are available in a dedicated GitHub directory\footnote{\url{https://github.com/ysBach/KMTNet_Toutatis}} which contains five notebooks:

\begin{itemize}
  \item \texttt{01\_data\_reduction}: the data reduction process described in Section \ref{sec: observation} and the generation of Table~\ref{table:obslog}.
  \item \texttt{02\_photometry\_comparison}: the comparison of our reduced data with \cite{1995Icar..117...71S} and internally developed Moving Object Detection Pipeline (MODP) to cross-check our photometry results.
  \item \texttt{03\_radiance\_factor}: the calculation of $ I/F $, $ f_1 $, $ f_2 $, and $ A_5 $ values from the photometry results and projected area calculation (see Appendix \ref{app: rotation}), and the generation of Figures \ref{fig:halpha} and \ref{fig:figradf}.
  \item \texttt{04\_cross\_section}: the calculation of $ S_\proj (\theta_\mathrm{bf},\, \phi_\mathrm{bf}) $ in Figure \ref{fig:figcrosssectionalarea} using \texttt{ParaView} software to check whether the $ S_\proj (t) $ values in Appendix \ref{app: rotation} (used IDL \texttt{POLYSHADE}) is reasonable.
  \item \texttt{05\_polarimetry\_plot}: the generation of Figure \ref{fig:grainsize} and a short discussion of the results in \cite{GeakeJE+DollfusA1986MNRAS}.
  \item \texttt{06\_KMTNet\_Seeing}: a short description about weather and seeing condition of our observations in Table~\ref{table:obslog}.
\end{itemize}

%%% BIBLIOGRAPHY %%%%%%%%%%%%%%%%%%%%%%%%%%%%%%%%%%%%%%%%%%%%%%%%%%%%

\end{document}